\documentclass[12pt]{article}
\usepackage[ansinew]{inputenc}
\usepackage{bm}
\usepackage{graphicx}
\newtheorem{propos}{Proposition}[section]
\newtheorem{teorem}{Theorem}[section]

\newtheorem{coro}{Corollary}[section]

\title{Spherically symmetric models for charged radiating
stars and voids: Theoretical approach}
\author{Francesc Fayos$^1$\thanks{Also at Laboratori de F\'{\i}sica Matem\`atica,
Societat Catalana de F\'{\i}sica, IEC, Barcelona.} ,
Jos\'e M.M. Senovilla$^2$ and Ram\'on Torres$^1$\thanks{Electronic address: labfm@hermes.ffn.ub.es}\\
$^1$ Departament de F\'\i sica Aplicada, UPC, Barcelona, Spain. \\
$^2$ Departamento de F\'{\i}sica
Te\'{o}rica, Universidad del Pa\'{\i}s Vasco,\\ Apartado 644,
48080 Bilbao, Spain.}
\date{}

\begin{document}
\maketitle

\begin{abstract}
We study the matching of a general spherically symmetric spacetime
with a Vaidya{-}Reissner{-}Nordstr\"{o}m solution. To that end, we study
the properties of spherically symmetric electromagnetic fields and develop the
proper {\em gravitational and electromagnetic} junction conditions. We prove
that generic spacetimes can be matched to a Vaidya{-}Reissner{-}Nordstr\"{o}m
solution or one of its specializations, and that these matchings have clear
physical interpretations. Furthermore, the non-spacelike nature of the
matching hypersurface is proved under very general hypotheses.
We obtain the fundamental result that any spherically symmetric body,
be it in evolution or not, has un upper limit for the total net electric
charge that carries.
\end{abstract}

PACS Nos: 04.20.Cv, 04.40.+c, 04.90.+e, 98.10.+z

\newpage

\section{Introduction}

Frequently, within the framework of General Relativity, in order
to describe new or interesting physical situations, or to
investigate certain theoretical aspects, the construction of a
model is required. This is commonly done by matching two given and
known space-times. Thus, for instance, models for stars (to study
the genesis of black holes, the possibility of escaping a black
hole region, the violation of the cosmic censorship conjecture,
etc.) and models for local inhomogeneities in a cosmological
context (to study the origin of galaxies in the primitive
Universe, the possibility of primordial black holes, the evolution
of the stars in the Cosmos, etc.) can be studied by means of the
junction of two different space-times.

With regard to the first type --construction of stellar models--
the common feature in most of them is the matching of two
space-times with special characteristics: one of them, which
describes the so called {\it interior} of the star; and another,
without matter but possibly with electromagnetic fields and/or null
radiation, which represents its {\it exterior}. Both regions are
separated by a timelike hypersurface, representing the
surface of the star. A very typical simplifying assumption is that the complete
space-time possesses spherical symmetry, so that in order to
describe the exterior of an object characterized only by its mass
the Schwarzschild vacuum solution is used; if, in addition, the
space-time possesses a net electrical charge, then the
Reissner{-}Nordstr\"{o}m (RN) solution is used; and if radial null radiation is
included then, instead of the previously cited metrics, the
solutions of Vaidya \cite{VM} and
Vaidya{-}Reissner{-}Nordstr\"{o}m (V{-}RN) \cite{krori}, are used,
respectively.

Thus, we can classify the works for stellar models according to
their {\it exterior}: The archetype with a Schwarzschild exterior
is the pioneering work by Oppenheimer-Snyder \cite{OS}, in which
the collapse of a dust cloud was studied. This work was later
generalized from two different points of view: on the one hand, by
considering a more general interior matter distribution. This was
done for instance by Misner and Sharp \cite{MSh}, where a perfect
fluid interior was considered, and by Bel and Hamoui \cite{Bel}
for the case of a fluid with anisotropic pressures. On the other
hand, by allowing for the existence of an electromagnetic field.
For example, when a charged dust interior with a
Reissner{-}Nordstr\"{o}m exterior was considered. Outstanding
examples of this treatment were the works by Novikov, de la Cruz
\& Israel, Markov \& Frolov, Hamoui, Vickers, Misra \& Srivastava,
Raychaudhuri and de Felice \& Maeda \cite{polvo}. Recently, Amos
Ori \cite{Ori} has found the explicit general solution for the
Einstein-Maxwell's equations in this case.

The next natural step for stellar models with a RN exterior was
to consider an even more general interior. This was initially achieved
by Bekenstein \cite{Bekenstein}, who considered a charged perfect
fluid interior, by using a generalization of the formalism (for
neutral perfect fluids) due to Misner and Sharp.
Some other interiors were later considered. This is the case of the
article by Florides \cite{Florides}, in which an exact solution
of the Einstein-Maxwell equations for a static, spherically
symmetric charge distribution is found. This represents an
extension, to the charged case, of a previous solution by Synge
\cite{Synge2}.

With regard to the radiating stars, that is, models with a Vaidya
exterior, we have to point out the works by Herrera {\it et al.}
(see \cite{HNrecopilacion} for a summary of their results) and
Santos {\it et al.} (see \cite{BOS} and references therein), where the
interior energy-momentum tensor is interpreted as a fluid with heat
flux. However, these works study only the shear-free
fluid case. Less restrictive is the article by Fayos {\it et al.}
\cite{fjlls2} where the preceding works on
Schwarzschild and Vaidya \cite{VM} interiors were generalized.

The complete case with a radiating and charged exterior (that is,
a V-RN exterior) has been treated by Oliveira and Santos
\cite{Oliveira}, but only in the case of a charged fluid interior
with heat flux but without shear. This article represents a
generalization of the work by Misner and Sharp, and also, of the
one by Bekenstein, in which dynamical equations for the spherical
collapse are given. Likewise, Herrera and N\'u\~nez
\cite{HNrecopilacion} considered a V-RN exterior, but only for a
charged shear-free perfect fluid interior, and with a flux of
non-polarized radial radiation. In their article they extend the
\textit{HJR method} --described in this same reference-- in order
to obtain the evolution of radiating charged fluid spheres
\cite{Medina}. They use a heuristic \textit{ansatz} which allows
them to integrate the Einstein-Maxwell equations taking into
account the matching conditions.

Regarding the construction of local inhomogeneities in a
cosmological context, the first fundamental work is the historically
relevant paper by Einstein and Strauss \cite{EinStr}, in which
``the influence of the expansion of space on the gravitation
fields surrounding the individual stars'' was studied. They considered a
vacuum Schwarzschild {\it interior} surrounded by a
Universe, modelized by a Robertson-Walker {\it exterior}. The
generalization, replacing the Schwarzschild solution with a Vaidya
interior, in order to describe the so called \textit{primordial
black holes}, can be found in \cite{Hacyan,Reed,Lake,fjlls}.
Lake and Hellaby \cite{LakeHellaby} used these models to study the
formation of naked singularities in the collapse of radiating
stars, as candidates for counterexamples of the cosmic censorship
conjecture.

It should be emphasized here that the actual matching procedure for the
Einstein-Straus and Oppenheimer-Snyder models are {\em mathematically
identical} and indistinguishable, because the two glued spacetimes are
exactly the same: vacuum Schwarzschild with dust Robertson-Walker. They
are in fact an outstanding example of {\em complementary matchings}, see
\cite{PSR}. This is in fact a general feature of the matching procedure and
thus we can always attack the two problems (description of stars, and modelization
of voids) in a common and unique framework. This is one of our aims in this
paper, and to that end we will follow the
philosophy put forward in \cite{PSR} which will allow us to treat
the two possibilities jointly in a natural way.

As a second goal of our paper, we wish to treat the fully general case in which
the matter content in the interior for stars (or the exterior for voids) is
left unrestricted, and the corresponding exterior (resp.\ interior) can contain
both radiation and electromagnetic fields. We will attack this problem by
studying the matching of two spherically symmetric space-times $\cal V$ and
$\bar{\cal V}$ through a time-like matching hypersurface $\Sigma$ in such a way
that one of the space-times will be completely general ($\cal V$), and
the other ($\bar{\cal V}$) may contain
an electromagnetic field and/or radial null radiation. This means that
$\bar{\cal V}$ will be described by a V-RN solution or one of its special
cases. We remark that our results will be general and
independent of the different combinations that might exist in
$\bar{\cal V}$: vacuum, only electromagnetic field, only radial
null radiation, or radiation and electromagnetic field.

The possible physical justification for including electromagnetic charge in
the models arises from several different considerations: i) first of all,
in the formation of a stellar object the repulsive
Coulomb force acting in every charged particle of the same sign of
the net charge of the star cannot be arbitrarily big, as is obvious. A simple
calculation, taking into account the opposing forces acting inside
a star with $n_b$ baryons, indicates that the net charge $Q$ must obey
the inequality $Q<10^{-36} n_b$ \cite{Glendenning}. This means that only
a small value of the net charge per nucleus is permitted,
but not necessarily zero; ii)
the existence of an electromagnetic field has drastic
consequences on the global space-time structure --compare, for
instance, the Schwarzschild solution with the Reissner{-
}Nordstr\"{o}m solution--. Thus, the evolution of a star with a net
electric charge $Q$ could be totally different to the evolution of
a neutral star, no matter how small $Q$ may be. New possibilities
--like that of a star avoiding the collapse towards a
singularity-- have been investigated since the appearance of the
previously cited article by de la Cruz and Israel, in which it was
shown for the first time that a charged object, after collapsing
beyond the event horizon, can expand and
re-emerge in another asymptotically flat region. Of course, all this depends on
the stability of the Cauchy horizons, see e.g \cite{Cauchy} and references
therein, a problem which is still subject to controversy;
finally, even if there is no electromagnetic field in $\bar{\cal V}$,
we could allow for a radial charge distribution in ${\cal V}$ with total net
charge vanishing. This is necessary to describe charge redistributions
that may take place in the interior of a star. A phenomenon like this happens, for
instance, in the \textit{hybrid stars}, in which there is global, but not
local, charge neutrality, see \cite{Glendenning}.

The plan of the paper is as follows. First, we will study, in
section \ref{capsimesf}, the properties of general spherically
symmetric space-times. In particular we write down the necessary
and sufficient conditions which guarantee the absence of curvature
singularities. Since we want our models to have a physical
meaning, we impose the fulfillment of the dominant energy
conditions and, as a first consequence of this, we obtain the
conditions such the \textit{mass function} will be non-negative.
In section \ref{seccem} we analyze the electromagnetic fields,
both null and non-null, that are compatible with the spherical
symmetry, and the restrictions on the distribution of charge that
arise at this level. Then we devote section \ref{capvrn} to the
study of the spherically symmetric V-RN space-time $\bar{\cal V}$.
In section \ref{capenlaces}, the gravitational {\em and
electromagnetic} matching conditions are derived. We distinguish
the cases with or without a surface charge current density in the
matching hypersurface. At the end of this section we study the
matching results, their interpretations and some consequences.
Next, in section \ref{seccont}, we analyze the restrictions on the
matter contents of the space-time $\cal V$ derived from the
matching conditions. In order to guarantee the physical meaning of
$\cal V$ --no matter which interpretation of the matter content is
used--, in section \ref{secemc} we will demand the fulfillment of
the dominant energy conditions to the energy-momentum tensor, and
we will examine the physical consequences that such constraints
imply. We end up with some conclusions, in which the main results
are summarized: they are the existence of a maximum for the total
charge of the bodies, the matchability of generic spacetimes to
V-RN or its specializations --these matchings having a clear
physical interpretation---, and the causality of the matching
hypersurface in general.

\section{Basics on spherically symmetric space-times.}
\label{capsimesf}

Let us consider a four-dimensional spherically symmetric
space-time $\cal V$, so that its line-element
can be expressed in radiative coordinates \cite{Bondi}
$\{x^{\mu}\}=\{u,R,\theta,\varphi\}$ ($\mu=0,1,2,3$) as
\begin{equation}
ds^2=-e^{4\beta}\chi du^2+2 \varepsilon e^{2\beta}\ du\ dR+R^2\
d\Omega^2 , \label{mI}
\end{equation}
where $\chi \equiv 1-2 m/R$, $\varepsilon^2=1$, $\beta$
and $m$ depend on $\{u,R\}$, and \mbox{$d\Omega^2\equiv
d\theta^2+\sin^2\theta d\varphi^2$}. It is easily checked that
$m(u,R)$ is the well-known mass function defined by Cahill \&
McVittie \cite{CMc}, which represents the total energy inside the
two-spheres with constant values of $u$ and $R$, see e.g.
\cite{Hay}.

The global structure and general properties of the space-time
depend on the behavior of the functions $m(u,R)$ and $\beta
(u,R)$. In this sense, for example, it can be seen that no
spherical closed trapped surfaces (see \cite{HE} for definitions)
are present in the whole spacetime if and only if $\chi$ is
non-negative everywhere (that is to say, $2m(u,R)\leq R$). In
fact, the \textit{two-spheres} defined by $u=u_c=$const.\ and
$R=R_c=$const.\ are closed trapped surfaces if and only if $\chi
(u_c,R_c)<0$, and thus the hypersurface $\chi =0$, which is the
boundary between regions with and without closed trapped spheres,
is the {\it apparent horizon}.

Similarly, all curvature invariants will be finite at $R=0$
\cite{Torres} if and only if (comma denotes partial derivative)
\begin{displaymath}
\beta_{,R}(0)= 0 \ ;\hspace{3mm} m(0)= m_{,R}(0)= m_{,RR}(0)= 0
\end{displaymath}
so that we have
\begin{propos}
The necessary and sufficient conditions preventing the existence
of a curvature singularity at $R=0$ are
\begin{eqnarray}
\lim_{R\rightarrow 0}\frac{\beta (u,R)- \beta _0 (u)}{R^2}&=&\beta _2(u),
\nonumber \\
\lim_{R\rightarrow 0}\frac{m(u,R)}{R^3}&=& m_3 (u),  \label{ro}
\end{eqnarray}
where $\beta _0 (u)\equiv \lim_{R\rightarrow 0}\beta (u,R)$, $\beta _2(u)$
and $m_3 (u)$ are finite functions of $u$.
\end{propos}
We shall later use these conditions.

In order to build up a model, some knowledge or constraints on the
energy-momentum content, evaluated over the region of the
space-time under study, are compulsory. The standard \textit{energy conditions},
see e.g. \cite{HE}, are among these physical constraints, so that
in this paper we will always demand that the
space-times under consideration fulfill the {\it dominant} energy
condition (DEC), which requires that the energy-flux 4-vector is
always non-spacelike \cite{HE}. For spherically symmetric
space-times with the metric expressed in radiative coordinates,
the DEC reduces to the following inequalities:
\begin{eqnarray}
\beta,_R &\geq& 0 ,\label{ec1o}\\
m,_R-R{\chi} \beta,_R&\geq& 0 , \label{ec2o}\\
{\chi} ^2 \beta,_R +\varepsilon \frac{2 e^{-2\beta}}{R} m,_u
&\geq& 0, \label{ec3o}\\
\lambda&\geq& |P_2| ,\label{ec4o}
\end{eqnarray}
where
\begin{eqnarray}
\lambda &=& -\frac{2{\chi}}{R} \beta,_R +\frac{2}{R^2}m,_R
+\frac{2}{R} \left({\chi} ^2\beta,_R^2+\varepsilon \frac{2 e^{-2
\beta}}{R} m,_u \beta,_R\right)^{1/2} ,\nonumber \\ P_2 &=&
(\frac{3}{R} -\frac{{\chi}}{R}+ 4{\chi} \beta,_R-\frac{6}{R}
m,_R)\beta,_R + 2{\chi} \beta,_{RR} -\frac{1}{R}
m,_{RR}+\varepsilon 2 e^{-2\beta} \beta,_{u R}.\nonumber
\end{eqnarray}

The inequality (\ref{ec2o}) can be rewritten as
\begin{eqnarray*}
( m e^{2 \beta})_{,R} \geq \frac{R}{2} (e^{2 \beta})_{,R}\geq 0
\end{eqnarray*}
where the last inequality follows from (\ref{ec1o}). Take an
arbitrary null hypersurface $u=u_c$=const. By integrating the
above inequality in such a hypersurface between $R_{1}$ and
$R_{2}\ (>R_{1})$ we obtain
\begin{eqnarray*}
m(u_c, R_{2}) e^{2 \beta(u_c, R_{2})} - m(u_c, R_{1}) e^{2
\beta(u_c, R_{1})} \geq \int^{R_{2}}_{R_{1}}
\frac{R}{2} (e^{2 \beta(u_c, R)})_{,R}\, dR \geq 0
\end{eqnarray*}
so that
\begin{equation}
m(u_c, R_{2}) e^{2 \beta(u_c, R_{2})} \geq m(u_c, R_{1}) e^{2
\beta(u_c, R_{1})} \label{II}
\end{equation}
 From here we deduce that if $m(u_{c},R_{1})\geq 0$, then
$m(u_{c},R)\geq 0$ for all $R\geq R_{1}$. In particular, it is
sufficient that $\lim_{R\rightarrow 0} m(u_c, R) e^{2\beta(u_c,
R)} \geq 0$ for the mass function to be non-negative on the whole
hypersurface $u=u_{c}$. Taking into account that $u=u_c$ is
arbitrary, we have proven the following
\begin{propos}\label{propm>0}
If a spherically symmetric space-time is such that the mass
function $m(u,R)$ is non-negative at $R=0$ and the dominant energy
conditions are fulfilled, then $m(u,R)\geq 0$ for all $R>0$.
\end{propos}
Actually, not all DEC's are needed here, but just conditions
(\ref{ec1o}-\ref{ec2o}). An interesting case for this proposition,
taking into account (\ref{ro}), is the following (compare with the
more restrictive \textit{prop. 6} in \cite{Hay})
\begin{coro}
If a spherically symmetric space-time has no curvature
singularity at $R=0$ and the DEC are fulfilled,
then the mass function $m(u,R)$ is non-negative everywhere.
\end{coro}

\section{Electromagnetic fields with spherical symmetry}
\label{seccem}

Starting with a spherically symmetric space-time
in its form (\ref{mI}) we can compute, through Einstein's
equations, its energy-momentum tensor $\mathbf{T}$. Choosing
an orthonormal cobasis $\{\mathbf{V}^{(\alpha)}
\}=\{{\bf u},{\bf n},\mbox{\boldmath{$\omega$}}_\theta,
\mbox{\boldmath{$\omega$}}_\varphi\}$ such that
\mbox{\boldmath{$\omega$}}$_\theta \equiv R\ d\theta$,
\mbox{\boldmath{$\omega$}}$_{\varphi}\equiv R\ sin\theta \
d{\varphi}$, $ {\bf n} $ is a space-like 1-form orthogonal to
\mbox{\boldmath{$\omega$}}$_\theta$ and
\mbox{\boldmath{$\omega$}}$_\varphi$, but otherwise {\it arbitrary}, and
${\bf u}$ is a time-like 1-form which completes the orthonormal
tetrad, one gets:
\begin{eqnarray}
\mathbf{T} &=& A\ \mathbf{u} \otimes \mathbf{u} + B\ \mathbf{n}
\otimes \mathbf{n}+ C\ (\mathbf{u} \otimes \mathbf{n} + \mathbf{n}
\otimes \mathbf{u})+\nonumber \\ && + D\ ( \bm{\omega}_\theta
\otimes \bm{\omega}_\theta+ \bm{\omega}_\varphi \otimes
\bm{\omega}_\varphi) \label{temes}
\end{eqnarray}
where $A,B,C$ and $D$ are functions of $u$ and $R$. This
energy-momentum tensor will be, in general, a mixture of fluids,
gas, radiation and perhaps an electromagnetic field. For the sake
of generality, we shall not identify all the particular fields and
components of the matter contents. Nevertheless, avoiding this
general rule, we shall in fact identify the electromagnetic part
of $\mathbf{T}$. This will reveal very useful because, among other
things, it provides junction conditions for the electromagnetic
field itself, independent of the gravitational ones, a fact which
plays a central role in the physics of the problem. Thus,
throughout this paper we will speak of the spherically symmetric
electromagnetic field ${\bf F}$, which is of course a 2-form
solution of the Maxwell equations and must be considered as a
partial source of the gravitational field so that
\begin{equation}
\mathbf{T}=\mathbf{P}+\mathbf{E} \label{PE}
\end{equation}
where $\mathbf{E}$ is the part associated to the electromagnetic
field
\begin{equation}
\mathbf{E}=\frac{1}{4 \pi}(F_{\alpha\gamma} F_\beta{}^\gamma
-\frac{1}{4} F_{\gamma\delta} F^{\gamma\delta} \eta_{\alpha\beta})\
\mathbf{V}^\alpha\otimes \mathbf{V}^\beta \label{temg}
\end{equation}
and $\mathbf{P}$ is the \textit{non-electromagnetic} part.
Obviously the case without electromagnetic field is trivially
included here when ${\bf F}=0$.
Assuming that $\mathbf{P}$ also has the structure
(\ref{temes}), then $\mathbf{E}$ will share this same structure,
so that $\mathbf{T}$, $\mathbf{P}$ and $\mathbf{E}$ will all be
invariant under the action of the SO(3) group.

Consider a general 2-form $\mathbf{F}$ and
demand that its energy-momentum tensor (\ref{temg}) has the
structure (\ref{temes}). This imposes several restrictions
on its components and one can easily prove that only two
possibilities arise, depending on whether $\mathbf{F}$ is null or not:
\begin{eqnarray}
{\bf F}= \psi  \ {\bf u} \wedge {\bf n} + \phi \
\mbox{\boldmath{$\omega$}}_\theta \wedge
\mbox{\boldmath{$\omega$}}_\varphi =
\Phi \left(\cos \eta \ {\bf u} \wedge {\bf n} \ +
\sin \eta \ \mbox{\boldmath{$\omega$}}_\theta \wedge
\mbox{\boldmath{$\omega$}}_\varphi\right), &\mbox{(non-null)}& \label{FIN}\\
\mathbf{F}= \Phi \ [\cos\eta (\mathbf{l} \wedge
\bm{\omega}_\theta)+ \sin\eta \ (\mathbf{l} \wedge
\bm{\omega}_\varphi )] \hspace{25mm} &\mbox{(null)}& \label{FSIN}
\end{eqnarray}
where $\Phi=\Phi (u,R)$ and $\eta=\eta (u,R,\theta,\varphi)$,
and $\mathbf{l}$ is a future-pointing
radial null vector field tangent to the (outgoing or ingoing) radial null
geodesics. We also assume that $\mathbf{l}$ is affinely parametrized.
The corresponding energy-momentum tensors $\mathbf{E}$ read
\begin{eqnarray}
\mathbf{E}= \frac{\Phi^2}{8 \pi}(\mathbf{u} \otimes \mathbf{u} -
\mathbf{n} \otimes \mathbf{n}+ \bm{\omega}_\theta \otimes
\bm{\omega}_\theta+ \bm{\omega}_\varphi \otimes
\bm{\omega}_\varphi), &\mbox{(non-null)}& \label{TemR}\\
\mathbf{E}= \frac{\Phi^2}{4\pi}\ \mathbf{l} \otimes \mathbf{l} \ .
\hspace{25mm} &\mbox{(null)}& \label{radem}
\end{eqnarray}

The previous restrictions are purely algebraic. In addition,
the electromagnetic field $\mathbf{F}$ must also satisfy the
Maxwell equations
\begin{eqnarray}
d \mathbf{F} &=&0 \label{EMdF}\\
\delta \mathbf{F} &=& 4\pi \mathbf{J} \label{EMd*F}
\end{eqnarray}
where $\delta$ is the co-differential (or divergence operator)
and $\mathbf{J}$ is the {\it electromagnetic current}
1-form. These equations impose further restrictions
on the components of $\mathbf{F}$ as follows:

a) {\it Non-null electromagnetic field}. Equations
(\ref{EMdF}-\ref{EMd*F}) applied to the electromagnetic field
(\ref{FIN}) lead to
\begin{eqnarray}
&&d( R^2 \phi) = 0 \ \ \ \ \ \ \Rightarrow \ \ \ \ \ \ \ \fbox{$\phi=
c/R^2$}\label{phires}\\
&&\frac{-1}{4 \pi R^2 } [ i(\mathbf{u}) d(R^2 \psi) \mathbf{n}+
i(\mathbf{n}) d(R^2 \psi) \mathbf{u}] =\mathbf{J} \label{Jres}
\end{eqnarray}
where $c$ is a constant\footnote{The possibility $c\neq 0$ will be
constrained later in this section and will turn out to be
impossible in our models as proved in section \ref{capenlaces}.}, and $i()$
indicates inner contraction. This implies that $\eta$
(and therefore $\phi$ and $\psi$) can only depend on $u$ and $R$.

Notice that $\mathbf{J}$ can be split into two components, one
along $\mathbf{u}$ and the other along $\mathbf{n}$. Thus, an
observer whose world lines are tangent to $\mathbf{u}$ will
measure a \emph{charge density} given by
\begin{displaymath}
\varrho=- J^\alpha u_\alpha= \frac{i(\mathbf{n}) d(R^2 \psi)}{4 \pi R^2}
\end{displaymath}
and a \textit{conduction current} \cite{BO}
\begin{equation}
\mathbf{j}\equiv \mathbf{J}-\varrho \mathbf{u}= -
\frac{i(\mathbf{u}) d(R^2 \psi)}{4 \pi R^2} \mathbf{n}\ .
\label{defjcond}
\end{equation}
The total charge $q$ within
a 2-sphere $S$ defined by constant values of $u$ and $R$
is \cite{Gravitation}:
\begin{equation}
q= \frac{1}{4 \pi} \int_{S} \mathbf{*F}
\hspace{0.5cm} \Longrightarrow \hspace{0.5cm} q=\psi R^2
\label{Qdef}
\end{equation}
where $\mathbf{*F}$ is the Hodge dual of $\mathbf{F}$.

b) {\it Null electromagnetic field}: The Maxwell
equations (\ref{EMdF}-\ref{EMd*F}) applied to the electromagnetic field
(\ref{FSIN}) provide now
\begin{eqnarray}
i(\mathbf{k}) d(R \Phi \cos\eta) = 0\ &,& \label{ems1}\\
i(\mathbf{k}) d(R \Phi \sin\eta) = 0\ &,& \label{ems2}\\
i(\bm{\omega}_\varphi) d(\cos\eta)\, \sin
\theta-i(\bm{\omega}_\theta) d(\sin\eta \sin\theta)= 0\ &,& \label{ems3} \\
 \frac{\Phi}{4 \pi} [  i(\bm{\omega}_\theta)
d(\cos\eta \sin\theta)\, \csc\theta- i(\bm{\omega}_\varphi)d(
\sin\eta)]\ \ \mathbf{l}\ =\mathbf{J} &,& \label{jsingular}
\end{eqnarray}
where $\mathbf{k} \cdot \mathbf{k}=0,\ \mathbf{l} \cdot
\mathbf{k}=-1$ and, in order to write $\mathbf{J}$ in the form
(\ref{jsingular}), we have used equations (\ref{ems1}-\ref{ems2}).
 From here we infer that the electromagnetic 4-current \ref{jsingular}
can only be null or zero and that the charge in any 2-sphere must be zero:
$q=0, \ \forall (u, R)$. Observe that if $\sin\eta \neq 0$ we can
write (\ref{jsingular}), using (\ref{ems3}), as
\begin{equation}
\mathbf{J} = -\frac{\Phi}{4 \pi } \csc\eta\ i(\bm{\omega}_\theta)
 d\eta\ \mathbf{l} \label{jsing1}
\end{equation}
while if $\sin\eta = 0$ then
(\ref{jsingular}) becomes
\begin{equation}
\mathbf{J} = \frac{\Phi}{4 \pi R} \cot\theta\ \mathbf{l}\ . \label{jsing2}
\end{equation}
The case of physical relevance is that with no sources for
the null electromagnetic field or, in other words, with $\mathbf{J}=0$.
 From (\ref{jsing1}-\ref{jsing2}), we see that this is only possible if
$i(\bm{\omega}_\theta) d\eta=0$ and $\sin\eta \neq 0$. However,
(\ref{ems3}) implies then
that $\eta,_\varphi=\cos\theta$, which is clearly incompatible
with the previous formulas. Therefore, we conclude that\emph{
there cannot exist a sourceless null electromagnetic field whose
energy-momentum tensor is $SO(3)$ invariant}. If these sources
existed the associated electromagnetic 4-current should be a
null vector, a possibility which has a questionable
physical interpretation (see e.g.\ \cite{Carmeli}, \cite{Carot} and
references therein). Thus, let us remark the well-known result that the only
electromagnetic field invariant under the action of the group
$SO(3)$ is the non-null one defined by
(\ref{FIN},\ref{phires}-\ref{Jres}).

As is well known, the tensor (\ref{TemR}) obeys the dominant
energy conditions by itself, and we have assumed that the whole
energy-momentum tensor (\ref{PE}) so does. Nevertheless, this does
not imply that $\mathbf{P}$ must satisfy the DEC. However, as we
wish to avoid negative energy densities or space-like energy
fluxes associated with $\mathbf{P}$, in what follows we are going
to demand the fulfillment of the dominant energy conditions for
this part too. Taking into account (\ref{TemR}), a straightforward
calculation proves that conditions (\ref{ec1o}) and (\ref{ec3o})
remain the same while the inequalities (\ref{ec2o}) and
(\ref{ec4o}) must be replaced by the stronger requirements:
\begin{eqnarray}
\Phi^2 &\leq& \frac{2}{R^2}(m,_R-R{\chi} \beta,_R) , \label{ec2}\\
2\Phi^2 &\leq& \lambda + P_2 ,\label{ec4}\\
0 &\leq& \lambda - P_2 .\label{ec5}
\end{eqnarray}

At this stage, we can already deduce important consequences from these
inequalities. For instance, the energy
condition (\ref{ec2}) has an important physical interpretation.
Using that $\Phi^2 =\phi^2 + \psi^2$, the definition (\ref{Qdef}) of
$q$, and the result (\ref{phires}) for
$\phi$, we can rewrite (\ref{ec2}) as:
\begin{equation}
q^2+c^2 \leq 2 R^2(m,_R-R{\chi} \beta,_R) .  \label{qm}
\end{equation}
Recall that $q$ is a function of $u$ and $R$. Thus, if the
space-time is given (that is, if we know $m(u,R)$ and
$\beta(u,R)$ explicitly), then this relation
can be interpreted as providing a maximum value to the
charge enclosed in every 2-sphere.

If we now demand also the absence of curvature singularities at $R=0$
(see (\ref{ro})), from (\ref{qm}) we get that, in a neighborhood
of $R=0$,
\begin{equation}
q^2+c^2 \leq 2 [3 m_3(u)-2 \beta_2(u)] R^4 + \mbox{higher order terms.}
\label{qcec}
\end{equation}
Hence, as $c$ is a constant and the second member of the
inequality is zero for $R=0$, we get that $c=0$ and $q(u,0)=0$,
and by using (\ref{phires}), we arrive at the following
\begin{propos}
\label{curvsingdec}
If a spherically symmetric space-time has no
curvature singularities at $R=0$ and the
energy condition (\ref{ec2}) is fulfilled, then
\begin{equation}
\phi=0\hspace{0.2cm},\hspace{0.4cm} q(u,R\rightarrow 0)=0.
\end{equation}
\end{propos}
In other words, the radial observers will not measure any
magnetic field (since its value is proportional to $\phi$) and
there is no charge contained in $R=0$.
In this case, (\ref{qm}) can be expressed
as\footnote{(\ref{ec2}) guarantees that the square root is real.}
\begin{eqnarray*}
\lim_{R\rightarrow 0}\frac{|q(u,R)|}{R^2}\leq
\sqrt{ 6 m_3(u)-4 \beta_2 (u)}\, \,  .
\end{eqnarray*}
As a matter of fact, the requirement of absence of singularities at $R=0$
is not necessary and, as is clear from the previous reasonings,
one only needs the fulfillment of
\begin{equation}
\lim_{R\to 0} R^2 (m,_R-R{\chi} \beta,_R) =0 . \label{refincq}
\end{equation}
Thus, there is a stronger version of proposition \ref{curvsingdec} where
the demand of absence of singularities is replaced by the milder
condition (\ref{refincq}), in which case there may be
a curvature singularity. As a particular case of this,
suppose that we can carry out
Taylor's expansions of $m$ and $\beta$ in a neighborhood of $R=0$:
\begin{eqnarray}
m(u,R)&=& m_0(u) +m_1(u) R+ m_2(u) R^2+ O(R^3),\nonumber\\
\beta(u,R)&=& \beta_0(u)+\beta_1(u) R+\beta_2(u) R^2+
O(R^3),\nonumber
\end{eqnarray}
then
\begin{equation}
\lim_{R\rightarrow 0}\frac{|q(u,R)|}{R}\leq
\sqrt{2 m_1(u)+4 m_0(u) \beta_1(u)}
\end{equation}
where, as before, the righthand side
is real due to (\ref{ec2}).

\section{The Vaidya-Reissner-Nordstr\"{o}m spacetime}\label{capvrn}
As stated in the introduction, the spherically symmetric
space-time $\bar{\cal{V}}$ will be taken as the
Vaidya-Reissner-Nordstr\"{o}m (V-RN) solution \cite{VM}, whose
line-element is given by (\ref{mI}) with
$2m(u,R)=2M(u)-Q^2/R$ and $\beta (u,R)=0$, that is
\begin{equation}
ds^2=-\left(1-\frac{2M(u)}{R}+\frac{Q^2}{R^2}\right) du^2+
2 \varepsilon\ du\ dR+R^2\ d\Omega^2\ ,
\label{VRNm}
\end{equation}
where $M(u)$ depends only on $u$ and $Q$ is a constant. Among the most
remarkable known particular cases of (\ref{VRNm}) we can cite:
if $M(u)$ is constant and $Q \neq 0$,
then it coincides with Reissner-Nordstr\"{o}m's (RN's) solution,
including its particular cases of Schwarzschild's solution ($Q=0$)
and flat Minkowski's spacetime ($M=Q=0$),
all of them very well-known and studied (see, e.g., \cite{HE});
if $M(u)$ is not constant and $Q=0$, then (\ref{VRNm}) reduces to
Vaidya's radiating solution \cite{VM}, whose structure and possible
extensions have been treated in
\cite{complete,I2,FMS}, including the possibility of
producing naked singularities (see, e.g., \cite{Joshi}.) A
general study of this solution was also carried out in
\cite{Torres}.

Whenever $Q\neq 0$, the 2-form
\begin{equation}
\mathbf{F}= \frac{Q}{R^2}\ \mathbf{u} \wedge \mathbf{n}\ .
\label{fb2}
\end{equation}
satisfies Maxwell's equations in the metric (\ref{VRNm}) and thus
$Q$ is interpreted as a charge and (\ref{fb2}) as an
electromagnetic field present in the spacetime.
The energy-momentum tensor ${\bf T}$ associated with (\ref{VRNm})
is given by
\begin{equation}
\mathbf{T} = \frac{ \varepsilon}{4 \pi R^2} \frac{dM(u)}{du}
\mathbf{l}\otimes \mathbf{l} + \mathbf{E} \label{TEX}
\end{equation}
where the electromagnetic part is given by (\ref{TemR}) with
$\Phi =\psi = Q/R^2$, clearly describing
a pure spherically symmetric non-null electrostatic
field of total charge $Q$,
and $\mathbf{l}=- du$ is the future-pointing radial null vector
tangent to the radial null geodesics with $u=$constant.
Thus, the non-electromagnetic part of the energy-momentum tensor
describes a kind of incoherent radially outgoing (respectively ingoing)
radiation filling the exterior region of the
spherical body whose total mass-energy, given by $M(u)$,
decreases with the retarded
time $u$ for $\varepsilon=-1$ (resp. increases if
$\varepsilon=+1$) as a result of the isotropic radial emission
(resp. absorbtion).
By analogy with (\ref{radem}), one might think that this kind of
radial radiation can be in fact electromagnetic radiation. However,
we have proved in section \ref{seccem} that this is not possible, since
there can be no null electromagnetic field without sources
satisfying the Maxwell equations (\ref{ems1}-\ref{jsingular})
 in this space-time whose stress-energy tensor matches the form
of $\mathbf{T-E}$ in (\ref{TEX}), as should be expected \cite{Carmeli}.

Observe that the energy conditions for the radiative
part of the energy-momentum tensor hold if
\begin{equation}
\varepsilon \frac{dM(u)}{du} \geq 0 , \label{cevb}
\end{equation}
and that the V-RN spacetime is an example without the properties
considered at the end of the previous section, because it has a
strong curvature singularity at $R=0$, and the milder condition
(\ref{refincq}) is also clearly violated whenever $Q\neq 0$. The
global structure of this space-time if $Q\neq 0$ was analyzed in
\cite{FMS}. There appear three cases depending on whether or not
there exists $u_1$ such that $M^2(u_1) = Q^2$ and, if not, on
whether $M^2(u)$ is greater than $Q^2$ or not. We refer the reader
to \cite{FMS, Torres} for an account of the particular features in
each case and for the corresponding Penrose diagrams of the V-RN
spacetimes.

\section{The gravitational-electromagnetic matching}\label{capenlaces}

In order to match a general spherically symmetric space-time
${\cal V}$ with the Vaidya-Reissner-Nordstr\"om space-time
$\bar{\cal V}$ \footnote{We will set an overbar on all variables,
parameters and functions of the space-time $\bar{\cal V}$ to
distinguish them from the corresponding objects of the general space-time
${\cal V}$.} across a time-like hypersurface $\Sigma$, we are going
to impose the gravitational {\it and} electromagnetic junction
conditions. Specifically, we demand that the Einstein
equations hold, in the distributional sense,
in the whole space-time in such a way that no
infinite jumps are allowed for the stress-energy tensor (thus, the
possibility of a thin shell of matter on $\Sigma$ is not considered).
On the other hand, the Maxwell equations are also assumed to be valid,
in a distributional sense, in the whole spacetime but allowing in this case
for the existence of a charge surface density so that the
electromagnetic 4-current may have an infinite jump at $\Sigma$.

The practical technique to perform the matching is to consider ${\cal V}$
and $\bar{\cal V}$ as divided, each one, by the candidate
matching hypersurfaces $\sigma$ (respectively, $\bar{\sigma}$)
into two parts --say $1$ and $2$ for ${\cal V}$ and $\bar{1}$ and
$\bar{2}$ for $\bar{\cal V}$--. If $\sigma$ and $\bar{\sigma}$ are
diffeomorphic, one can identify corresponding points in them and
then try to perform the gluing which, in principle, can be
done in four different ways, namely, $1-\bar{2}$; $1-\bar{1}$; $2-\bar{2}$;
$2-\bar{1}$. We can select one of them by choosing the relative
sign of the normal vectors to the
matching hypersurfaces $\sigma$, $\bar{\sigma}$, see \cite{PSR}
for details. Notice that, as these normal vectors are spacelike then we
can always choose the tetrad such that ${\bf n}$ and $\bar{{\bf n}}$
coincide, on $\sigma$ and $\bar{\sigma}$ respectively, with them.
Thus, from now on the unit normal vectors will also be denoted by
${\bf n}$ and $\bar{{\bf n}}$.

Now, once $\sigma$ and $\bar{\sigma}$ have been identified
in the new glued spacetime there is no need to distinguish them
and they will be simply termed as $\Sigma$ if there is no confusion.
If $\Sigma$ is timelike and preserves the spherical symmetry then
it can be described by intrinsic coordinates
$\{\xi, \vartheta,\varphi\}$ where $\xi$ is a timelike coordinate
and where $\{ u(\xi)$, $R(\xi)$, $\theta=\vartheta$,
$\phi=\varphi\}$ and $\{\bar{u}(\xi)$,
$\bar{R}(\xi)$, $\bar{\theta}=\vartheta$, $\bar{\phi}=\varphi\}$
are the parametric representations of $\sigma$ in $\cal{V}$ and
of $\bar{\sigma}$ in $\bar{\cal{V}}$, respectively.
Before going any further,
and as pointed out in \cite{ClarkDray,MarcJose}, one has to
specify how the tangent planes at every point $p\in \sigma$
and at its corresponding point $\bar{p}\in \bar{\sigma}$ must be
identified in order to construct a well-defined geometry in the
whole glued space-time. To do that, we proceed as follows:
firstly we identify the vector fields associated to the angular
variables, $\partial/\partial\theta$ with
$\partial/\partial\bar{\theta}$, and $\partial/\partial\varphi$
with $\partial/\partial \bar{\varphi}$, at $\Sigma$; secondly
we consider the two vector fields defined
on $\sigma$ and $\bar{\sigma}$ respectively by
\begin{eqnarray}
\left.\left(\frac{\partial{R(\xi) }}{\partial \xi} \frac{\partial}{\partial
R}+ \frac{\partial{u(\xi) }}{\partial \xi}
\frac{\partial}{\partial u}\right)\right\vert_{\sigma}
\hspace{.5cm} \mbox{and} \hspace{.5cm}
\left.\left(\frac{\partial{\bar{R}(\xi) }}{\partial \xi}
\frac{\partial}{\partial \bar{R}}+ \frac{\partial{\bar{u}(\xi)
}}{\partial \xi} \frac{\partial}{\partial
\bar{u}}\right)\right\vert_{\bar{\sigma}}\label{vectorsdef}
\end{eqnarray}
and identify them since they both represent $\partial/\partial
\xi$ at $\Sigma$; thirdly, take the unit normal vectors
$\mathbf{n}$ (resp., $\mathbf{\bar{n}}$) for $\sigma$
(resp., $\bar{\sigma}$) defined, except for a sign, by
being orthogonal to the three previous vectors at $\Sigma$, and
choose these signs in such a way that every curve crossing $\Sigma$
through a point $p\equiv\bar{p}$ must have a unique well-defined
tangent vector there \cite{PSR}; and fourthly, by
doing so, the relative sign $\epsilon_n$ ($\epsilon_n^2=1$)
of the normal unit vectors has been fixed
so that we can identify them thereby achieving
a complete identification of the
two tangent planes. As a by-product, this process
determines an orientation and an arrow of time for the glued
space-time if these concepts are well defined or fixed in some
way in ${\cal V}$ and $\bar{{\cal V}}$.

At this stage, we impose the Darmois gravitational junction
conditions \cite{Darmois}\cite{PSR}, which are the best suited for
our purposes, requiring that the first and second fundamental
forms of $\Sigma$ be identical when computed from either ${\cal
V}$ or $\bar{{\cal V}}$ . After the appropriate calculations,
these matching conditions can be written in two (equivalent)
complete sets, depending on the sign $\epsilon_n$, as follows

{\bf Case $\varepsilon \bar{\varepsilon}=\epsilon_n$}
\begin{eqnarray}
R &\stackrel{\Sigma}{=}&\bar{R} \ , \hspace{2cm} \label{jc1} \\
\varepsilon \dot u e^{2\beta}
&\stackrel{\Sigma}{=}&\bar{\varepsilon}
\dot{\bar{u}} , \hspace{2cm}\label{jc2} \\
m &\stackrel{\Sigma}{=}&\bar{M} - \frac{\bar{Q}^2}{2\bar{R}},
\hspace{1.5cm} \label{jc3} \\
-\varepsilon \left [ \left(1-\frac{2 m}{R}\right) {\beta},_R - \right. \left.
\frac{m,_R}{2R}\right] e^{2\beta} {\dot u}+{\beta},_R{\dot R}
&\stackrel{\Sigma}{=}& \bar{\varepsilon}
\frac{\bar{Q}^2}{4\bar{R}^3}{\dot{\bar{u}}}\ . \label{jc4}
\end{eqnarray}

{\bf Case $\varepsilon \bar{\varepsilon}=- \epsilon_n$}
\begin{eqnarray}
R &\stackrel{\Sigma}=&\bar{R} \ ,\label{mc1} \\ \varepsilon \chi
e^{2\beta} \dot u -2\dot{R} &\stackrel{\Sigma}=& -\bar{\varepsilon}
\bar{\chi} \dot{\bar{u}} , \label{mc2}\\ m &\stackrel{\Sigma}=&
\bar{M} -
\frac{\bar{Q}^2}{2\bar{R}},\hspace{1.5cm} \label{mc3}  \\
\varepsilon \left[  \chi  \left( \chi \beta,_R- \frac{m,_R}{2 R}
\right)+ \varepsilon  \frac{m,_u e^{-2\beta}}{R} \right] e^{2\beta}
\dot{u}+ \left( \frac{m,_R}{R}- \chi \beta,_R \right) \dot{R}
&\stackrel{\Sigma}{=}&
\bar{\varepsilon}\frac{\bar{Q}^2}{4\bar{R}^3}\bar{\chi}\dot{\bar{u}}\ .
\label{mc4}
\end{eqnarray}
Here $\stackrel{\Sigma}{=}$ means that we have to
compute both sides of the equality at $\Sigma$, and the overdots
stand for derivatives with respect to $\xi$.

In practice, we can understand the equivalence of the
two sets of matching conditions by noting that they are suitable to
describe the prolongation, when crossing $\Sigma$, of a given
radial null geodesic with $u=$constant by either a radial null geodesic
with $\bar{u}=$constant (if $\varepsilon \bar{\varepsilon}=\epsilon_n$),
or not (if $\varepsilon \bar{\varepsilon}=-\epsilon_n$),
see \cite{Torres} for more details.

The above takes care of the {\it gravitational} matching, but we still
require an appropriate junction of the electromagnetic fields. To that
end, we present now the {\em general} matching conditions for the
electromagnetic field (for arbitrary space-times) and then
we will first particularize them to the spherically symmetric case and in
a latter step to the case in which one of the matching space-times
($\bar{\cal V}$) is the V-RN solution.
Let $\mathbf{F}$ and $\bar{\mathbf{F}}$ be the respective electromagnetic
2-forms at $\cal V$ and $\bar{\cal V}$, so that they satisfy
the Maxwell equations
\begin{eqnarray}
d{\mathbf F}=0 \ \ & & \ \  \delta {\mathbf F}=4 \pi \mathbf{J}
\\ d\mathbf{ \bar F}=0 \ \ & & \ \ \delta \bar{\mathbf{F}}=4 \pi
 \mathbf{ \bar J}
\end{eqnarray}
on ${\cal V}$ and $\bar{\cal V}$, respectively.
The whole electromagnetic field is defined to be
\begin{equation}
\mathbf{\cal{F}}=\mathbf{F} (1-\theta_\Sigma)+\mathbf{\bar{F}}
\theta_\Sigma
\end{equation}
where the Heaviside theta function $\theta_\Sigma$ is defined by
$\theta_\Sigma |_{\cal V}=0$ and $\theta_\Sigma|_{\bar{\cal V}}=1$.
Thus, it is immediate to get
\begin{eqnarray}
d\cal F &=&(\mathbf{F} \wedge \mathbf{ n}-\mathbf{\bar{F}} \wedge
\mathbf{\bar n})\, \delta_\Sigma \label{difF}\\ \delta\cal F &=& 4
\pi [\mathbf{J} (1-\theta_\Sigma)+\bar{\mathbf{J}} \theta_\Sigma]+
[i({\bf n})\mathbf{F}-i({\bar\mathbf{n}})\mathbf{\bar{F}}]\,
\delta_\Sigma \label{difrF}
\end{eqnarray}
where $\wedge$ is the exterior product and $\delta_\Sigma$ is the
normalized Dirac delta with support on $\Sigma$, see e.g.
\cite{ClarkDray,MarcJose}, defined by $d\theta_\Sigma=\mathbf{n}\,
\delta_\Sigma$. It follows from (\ref{difF},\ref{difrF}) that the
fulfillment of the Maxwell equations for $\mathbf{\cal{F}}$ in a
distributional sense provides the sought-after {\em
electromagnetic} junction conditions:
\begin{eqnarray}
{\bf F} \wedge {\bf n} \stackrel{\Sigma}{=} {\bar{\bf F}} \wedge
{\bar{ \bf n}} \label{emc2}\\
\mathcal{J} \equiv \mathbf{J} (1-\theta_\Sigma)+\bar{\mathbf{J}}
\theta_\Sigma+ \mathcal{K} \delta_\Sigma,\label{nueva}
\end{eqnarray}
where $\mathcal{J}$ is the distributional 4-current and
\begin{equation}
\mathcal{K}\stackrel{\Sigma}{\equiv} \frac{1}{4 \pi} [i({\bf
n})\mathbf{F}-i({\bar\mathbf{n}})\mathbf{\bar{F}}]\label{nova}
\end{equation}
is the \emph{surface current density} 4-vector (obviously defined
only on $\Sigma$).

If we do not wish to allow the total 4-current to have
infinite jumps at $\Sigma$, then from (\ref{nueva}) and (\ref{nova})
we must impose
\begin{equation}
i({\bf n}){\bf F} \stackrel{\Sigma}{=} i({\bar{\bf n}}){\bar{\bf
F}} \label{emc1}
\end{equation}
so that, taking into account (\ref{emc2}) we derive
$\mathbf{F}\stackrel{\Sigma}{=}\bar{\mathbf{F}}$ {\em in this case}.

Let us now consider the particular case we are interested in:
spherical symmetry. From conditions (\ref{emc2}) one immediately
infers that a non-null electromagnetic field (resp., a null one)
can only be matched with another non-null field (resp., null).
But as we proved before, null electromagnetic fields are not compatible with
spherically symmetric energy-momentum tensors, so that only the non-null
case must be considered for our purposes\footnote{If one does not demand the
invariance of the null field energy-momentum tensor and computes the junction
conditions (\ref{emc2},\ref{nueva}) for the null-null case one easily gets
 $\mathcal{K}=0$ and $\mathbf{F} \stackrel{\Sigma}{=} \bar{\mathbf{F}}$,
see \cite{Torres}.}. Hence, the electromagnetic fields take the form
(\ref{FIN})
\begin{displaymath}
{\bf F}= \psi \ {\bf u} \wedge {\bf n} + \phi \
\mbox{\boldmath{$\omega$}}_\theta \wedge
\mbox{\boldmath{$\omega$}}_\varphi \hspace{0.5cm} \mbox{and}
\hspace{0.5cm} {\bar{\bf F}}= \bar{\psi} \ \bar{\bf u} \wedge
\bar{\bf n} + \bar{\phi} \ \bar{\bm{\omega}}_\theta \wedge
\bar{\bm{\omega}}_\varphi
\end{displaymath}
in $\cal V$ and $\bar{\cal V}$, respectively, so that
the matching conditions (\ref{emc2},\ref{nueva}) imply
\begin{eqnarray}
\phi \stackrel{\Sigma}{=} \bar{\phi}  \label{phiphi}\\
\mathcal{K}= \frac{1}{4 \pi}(\bar{\psi}-\psi) \mathbf{u}\ .\label{new}
\end{eqnarray}

Let us finally particularize to the matching with the V-RN
solution. Condition (\ref{phiphi}) applied to the
electromagnetic fields (\ref{FIN}) and (\ref{fb2}) becomes
\begin{equation}
\phi \stackrel{\Sigma}{=} 0 \label{phi}
\end{equation}
which, together with (\ref{phires}), leads to vanishing
of $\phi$ on the entire $\cal V$, and not just on $\Sigma$:
\begin{displaymath}
\phi = 0 . \label{phi0}
\end{displaymath}
This result means that the radially moving observers cannot measure
any magnetic fields on $\cal V$. Similarly,
taking into account that the ``charge function'' on $\cal V$
is given by \mbox{$q(u,R)=\psi R^2$} [see
(\ref{Qdef})], we can define the \textit{total charge}
$Q\stackrel{\Sigma}{\equiv} \psi R^2$. Using now (\ref{fb2}),
we can express $\mathcal{K}$ of (\ref{new}) as
\begin{displaymath}
\mathcal{K} \stackrel{\Sigma}= \frac{1}{4 \pi R^2}(\bar{Q}-Q)
\mathbf{u}
\end{displaymath}
so that, using (\ref{Qdef}), we are led to define the \emph{total
surface charge} in $\Sigma$ by
\begin{displaymath}
Q_\Sigma \equiv \bar{Q}-Q \stackrel{\Sigma}{=}\bar{Q}- \psi R^2 .
\end{displaymath}
This is a very natural result and implies that the sum of the
total charge $Q$ contained in $\cal V$ plus the total surface
charge $Q_\Sigma$ must be constant ($=\bar Q$). Of course, the
presence of infinite jumps on the electromagnetic 4-current is a
mathematical idealization that can, in many cases, simplify the
complexity of describing some models with a large concentration of
charge at $\Sigma$. However, the absence of infinite jumps in the
4-current is more realistic physically and, in fact, it is in
agreement with the experimental observations \cite{Jackson}.
Therefore, if in particular we do not allow for infinite jumps in
the 4-current or, in other words, if there is not a surface charge
we obtain
\begin{equation}
Q \stackrel{\Sigma}{\equiv} \psi R^2 \stackrel{\Sigma}{=} \bar{Q}.
\label{q=Q}
\end{equation}
Then the total charge $Q$ must also be constant.

In the rest of this section, we are going to focus on this
physically more realistic case when (\ref{q=Q}) is fulfilled.
Then, the complete set of matching equations
(\ref{jc1})-(\ref{jc4}), (\ref{phi}) and (\ref{q=Q}) (or alternatively
(\ref{mc1})-(\ref{mc4}), (\ref{phi}) and (\ref{q=Q})) are the
necessary and sufficient conditions for the matching of a general
spherically symmetric metric with the V-RN solution through a
general spherically symmetric timelike hypersurface, and they
provide relations between relevant quantities at both sides of $\Sigma$.
Nonetheless, it is remarkable that a linear combination of
(\ref{jc1}), (\ref{q=Q}) and (\ref{jc4}) leads to
\begin{equation}
\varepsilon \left[\left(1-\frac{2 m}{R}\right) {\beta},_R- \right.
\left. \frac{m,_R}{2 R}+\frac{Q^2}{4 R^3} \right] {\dot u} -
{\beta},_R e^{-2\beta}{\dot R}\stackrel{\Sigma}{=} 0 ,
\label{jc4bis}
\end{equation}
or alternatively, for the other case with $\varepsilon
\bar{\varepsilon}=-\epsilon_n$, one similarly finds
\begin{equation}
\varepsilon \left[ \chi \left( \chi \beta,_R - \frac{m,_R}{2 R}
\right) + \varepsilon \frac{m,_u e^{-2 \beta}}{R}+ \frac{Q^2
\chi}{4 R^3} \right] \dot{u} + \left( \frac{m,_R}{R} -\chi
\beta,_R-\frac{Q^2}{2 R^3} \right) e^{-2 \beta} \dot{R}
\stackrel{\Sigma}= 0 . \label{mc4bis}
\end{equation}
Clearly, these relations involve quantities of the space-time
${\cal V}$ {\em only}, but {\it not}\/ of $\bar{\cal V}$. This
means that equation (\ref{jc4bis}) (or (\ref{mc4bis})) is a
necessary condition that the space-time ${\cal V}$, by itself,
must fulfill on $\Sigma$ in order to be matchable to a V-RN
solution. The physical meaning of (\ref{jc4bis}) (or
(\ref{mc4bis})) generalizes the results for standard models in
which normal pressures must vanish on $\Sigma$ when matching to a
vacuum, as will be explained in section \ref{seccont} in detail.

Given the previous remarks, one can now derive much more information
from the matching equations. To extract this information, one
has to follow paths which depend on the known data of the problem
under consideration. A physically very interesting case arises when
$m(u,R)$ and $\beta(u,R)$ are given, so that the spacetime $\cal V$ is
completely known, and we want to ascertain whether or not $\cal V$
is {\em matchable} to a V-RN solution, if so {\em where}, and finally
{\em to which particular} V-RN metric, that is, for which particular
$\bar{M}(\bar{u})$ and  $\bar{Q}$. In order to solve this problem,
first of all we treat $\bar{Q}(\stackrel{\Sigma}=Q)$ either as known
or as a parameter
on which the final results will depend. This is a key point.
Then, we proceed as follows: for the $\varepsilon \bar{\varepsilon}=
\epsilon_n$ case (resp., $\varepsilon \bar{\varepsilon}=-
\epsilon_n$), the equation (\ref{jc4bis}) (resp., (\ref{mc4bis})),
can be considered as an ordinary differential equation for
$R(u)$, with the form $dR/du=F(u,R;\bar{Q})$, so that if $m(u,R)$
and $\beta (u,R)$ are such that $F(u,R;\bar{Q})$ satisfies Lipschitz's
conditions we can find the solution $R(u;c_1,\bar{Q})$, where $c_1$ is
an integration constant. If $R(u;c_1,\bar{Q})$ defines a time-like
$\Sigma$ on $\cal V$, then using (\ref{jc3}) we can determine
$\bar{M}\stackrel{\Sigma}{=}\bar{M}(u;c_1,\bar{Q})$.
Now, by integrating (\ref{jc2}) for the first case --or
(\ref{mc2}) for the second-- we get $\bar{u}(u;c_1,\bar{Q})$, except for
a new additive constant $c_2$,\footnote{The specific value of $c_2$ is
irrelevant since it only provides an origin for the {\it times}
$u$ and $\bar{u}$, and therefore we choose, for the sake of simplicity,
$c_2=0$.} while from (\ref{jc1}) (or, respectively, from (\ref{mc1}))
we get $\bar{R}(u;c_1,\bar{Q})$. These two functions define the
hypersurface $\Sigma$ in the space-time $\bar{\cal V}$.  Finally,
due to the fact that $\bar{M}(\bar{u})$ is a function of only $\bar{u}$,
by combining $\bar{M}(u;c_1,\bar{Q})$ with  $\bar{u}(u;c_1,\bar{Q})$
we get $\bar{M}(\bar{u};c_1,\bar{Q})$ and the problem is
completely solved.

The solutions $R(u;c_1,\bar{Q})$ give a {\em two-parameter} family
of matching hypersurfaces through which the space-time ${\cal V}$
is matchable with a V-RN solution. The different values for $c_1$
(and of course $\bar{Q}$) will simply lead to the different
particular V-RN space-times which match at the various $\Sigma$'s,
providing explicitly $\bar{M}(\bar{u})$ (and $\bar{Q}$). In order
to interpret the physical differences between these $\Sigma$'s,
consider their intersection with a given null hypersurface
$u=u_0=$constant. The values of $R\stackrel{\Sigma}{=}\bar{R}$ and
$\bar{M}$ at these intersections are denoted by
$R_0=R(u_0;c_1,\bar{Q})$ and $\bar{M}_0=\bar{M}(u_0;c_1,\bar{Q})$,
so that they depend on the value of $c_1$, that is, on the
particular $\Sigma$. Eliminating $c_1$ from $\bar{M}_0$ and $R_0$
we obtain $\bar{M}_0=\bar{M}_0(u_0,R_0,\bar{Q})$ which gives, at
the given $u_0$, the value of $\bar{M}_0$ as a function of $R_0$.
Use of (\ref{jc3}) [or (\ref{mc3})] leads to the explicit
expression
\begin{displaymath}
\bar{M}_0 (u_0,R_0,\bar{Q}) =  m(u_0,R_0) + \frac{\bar{Q}^2}{2 R_0}\ .
\end{displaymath}
Differentiating here with respect to $R_0$ and taking into account
the energy condition (\ref{ec2}) with $\bar{Q}\stackrel{\Sigma}{=}Q$
we get
\begin{displaymath}
\frac{d \bar{M}_0}{d R_0} =  m_{,R}(u_0,R_0) -
\frac{\bar{Q}^2}{2 R_0^2} \geq R_0
\left.\left(\chi \beta,_R\right)\right\vert_{u_0,R_0} \, .
\end{displaymath}
Finally, if the hypersurface at $u_0$ is not at a region with
trapped 2-spheres, that is to say, $\chi (u_0,R_0)\geq 0$, then
from the energy condition (\ref{ec1o}) we finally infer
\begin{displaymath}
\frac{d \bar{M_0}}{d R_0}\geq 0\, .
\end{displaymath}
In other words, given a space-time $\mathcal{V}$ take the
2-spheres (which belong to a matching hypersurface $\Sigma$)
defined by $\Sigma \cap \{u=u_0\} \cap \{\chi \geq 0\}$. These
will have an area proportional to $R_0^2$ and contain a total mass
$M_0$. The previous result tells us that \textit{smaller} such
2-spheres will contain non-bigger masses.

\section{Implications on the matter content of $\cal V$}\label{seccont}
In the previous sections we have not interpreted the matter
contents of the space-time $\cal V$ apart from assuming that it
may contain an electromagnetic field. Our purpose in this section
is to investigate some of the different interpretations for the
non-electromagnetic matter content and to elucidate the restrictions imposed
by the matching conditions, found in the previous section, on the
electromagnetic field as well as on the non-electromagnetic part
of the energy-momentum tensor.

Let us start with the electromagnetic field in ${\cal V}$,
given by (\ref{FIN}) with (\ref{Jres}) and
$$\psi\stackrel{\Sigma}=\bar{Q}/R^2,\,\,\, \phi=0 .$$
Taking the derivative of the first of these equations
along $\Sigma$ we get, on using (\ref{Jres})
\begin{equation}
i(\mathbf{u})d(R^2 \psi) \stackrel{\Sigma}=0\ \  \Longrightarrow\ \ \
\mathbf{J}\stackrel{\Sigma}=\frac{-1}{4 \pi R^2 } i(\mathbf{n})
d(R^2 \psi) \mathbf{u}. \label{jsigmaconduc}
\end{equation}
This tells us that $\mathbf{J}$ is a time-like vector on $\Sigma$,
furthermore comoving with $\Sigma$ at $\Sigma$. Outside $\Sigma$,
though, the form of $\mathbf{J}$ is unrestricted, and in general
it has two components, one along the direction of the time-like
$\mathbf{u}$, which is the convection current with respect to
$\mathbf{u}$ (assuming that $\mathbf{u}$ is chosen to be the
4-velocity of the charged fluid), and another in the direction of
$\mathbf{n}$, representing the corresponding conduction current
with respect to $\mathbf{u}$, which {\em necessarily} vanishes at
$\Sigma$. Therefore, any function $\psi$ that satisfies
(\ref{q=Q}) provides us with a regular electromagnetic field valid
for $\cal V$.\footnote{If we allowed for infinite jumps on ${\bf
J}$ the problem would be \textit{easier} since there would be no
constraints on $\psi$, not even on the matching hypersurface, so
that any function $\psi$ would provide  a regular electromagnetic
field valid for $\cal V$.}

Now we can ask ourselves the physically relevant question of whether
the electromagnetic field can be chosen only with
convention, or only with conduction, current. This is clearly
equivalent, according to (\ref{Jres}), to finding out the causal
character of the one-form $d(R^2\psi)$, because the sign of ${\bf
J}\cdot {\bf J}$ is opposite to that of $d(R^2 \psi )\cdot
d(R^2 \psi )$. Thus ${\bf J}$ is time-like (resp.\ space-like) and
thus can be aligned with ${\bf u}$ (resp.\ ${\bf n}$) if
$d(R^2\psi)$ is spacelike (resp.\ timelike). Observe that in the
second case $\mathbf{J} \stackrel{\Sigma}{=} 0$ necessarily.

Finally, the energy-momentum tensor $\bf E$ associated with the non-null
electromagnetic field in $\cal V$ is now (\ref{TemR}) with $\phi=0$:
\begin{equation}
\mathbf{E} = \frac{\psi^2}{8 \pi}(\mathbf{u}{\otimes}\mathbf{u} -
\mathbf{n}{\otimes}\mathbf{n} +
\bm{\omega}_{\theta}{\otimes}\bm{\omega}_{\theta} +
\bm{\omega}_{\varphi}{\otimes} \bm{\omega}_{\varphi}) . \label{EI}
\end{equation}

Let us now pass to the rest of the matter content. Hitherto, we
only know that its energy-momentum tensor, ${\bf P}={\bf T}-{\bf E}$,
takes a form of type (\ref{temes}).
As is known, see e.g. \cite{Synge}, the energy-momentum tensors for
different {\em physical types} of matter distributions can, in fact,
have precisely the same components.
Therefore ${\bf P}$ is opened to different interpretations,
depending on the physical processes we want to describe.

One of the simplest interpretations that we can put forward,
containing the minimum degrees of freedom, is the case of
a {\it charged non-perfect fluid embedded in a radially directed
null radiation}. The 4-velocity is taken to be
${\bf u}$ so that the fluid is comoving with
$\Sigma$ by construction. The total energy-momentum tensor for
this case is then written as
\begin{equation}
{\bf T} = (\mu + p)\ {\bf u} {\otimes} {\bf u} + p\ {\bf g} +
\Omega^2\ {\bf \ell} {\otimes} {\bf \ell} + {\bf \Pi} + {\bf E} ,
\label{TI}
\end{equation}
where $\mu$ is the energy density of the fluid,
$p$ its isotropic pressure, ${\bf \Pi}$ its anisotropic
pressure tensor (which is traceless and orthogonal to ${\bf u}$),
$\Omega^2$ is the null radiation energy density,
${\bf \ell} = {\bf u} + s\, {\bf n} \,\, (s^2=1)$ is a
radially directed null vector field, and ${\bf E}$ is given in
(\ref{EI}).

Obviously, the form (\ref{TI}) of the energy-momentum tensor
is not unique and another physically relevant possible interpretation, also
having the minimum degrees of freedom, is as a {\em charged non-perfect fluid
with heat conduction}. In this case we write
\begin{equation}
{\bf T} = (\tilde\mu + \tilde p){\bf u} {\otimes} {\bf u} + \tilde
p {\bf g} + {\mathbf h \otimes \mathbf u} + {\mathbf u \otimes
\mathbf h} +{\bf \tilde\Pi} + {\bf E} , \label{TI2}
\end{equation}
where $\mathbf{h}\propto \mathbf{n}$ is the heat flux vector and $\tilde\mu$,
$\tilde p$ and $\bf \tilde\Pi$ have the usual interpretation. Of course,
(\ref{TI},\ref{TI2}) are not the only possible cases and many other
physically realistic decompositions can be written according to the
particular situation.

In any case, and following the philosophy proposed in \cite{fjlls2}, for
every given $m(u,R), \beta (u,R)$ and $\psi (u,R)$ such that
(\ref{q=Q}) is satisfied and a timelike $\Sigma$ exists,
the velocity vector field $\mathbf{u}$ is fixed only on $\Sigma$
and, provided that it is time-like and orthogonal to
the 2-spheres $\{u,R\}=$constants, we have freedom to choose it in the
rest of $\cal V$. Once $\mathbf{u}$ is chosen  we
can compute, by means of (\ref{Jres}), the 4-current
$\mathbf{J}$, and by using Einstein's equations,
$\mu$, $p$, $\Omega^2$ and ${\bf \Pi}$ for (\ref{TI}), or
$\tilde\mu$, $\tilde p$, $\mathbf{h}$ and $\bf \tilde\Pi$ for
(\ref{TI2}), or the corresponding quantities for any other desired
interpretation of $\mathbf{P}$.

It is known that some relations between the energy-momentum tensor
at both sides of $\Sigma$ follow from the matching of two
space-times \cite{Israel,MarcJose}. They read
\begin{eqnarray}
{\bf T}_{\mu\nu} \tau^\mu n^\nu &\stackrel{\Sigma}{=} &\bar{\bf
T}_{\mu\nu} \bar{\tau}^\nu \bar{n}^\nu\ , \label{Tsigmact}\\ {\bf
T}_{\mu\nu} n^\mu n^\nu&\stackrel{\Sigma}{=} &\bar{\bf T}_{\mu\nu}
\bar{n}^\mu \bar{n}^\nu\ , \label{Tsigmacn}
\end{eqnarray}
where $\tau^\mu$ (equivalently $\bar{\tau}^\mu$) is {\em any}
vector tangent to $\Sigma$. Therefore, for the first interpretation
proposed in (\ref{TI}), the non-trivial relations
deriving from (\ref{Tsigmact},\ref{Tsigmacn}) are
\begin{eqnarray}
\mathbf{\ell} &\stackrel{\Sigma}{=}& \mathbf{l} , \\
\Omega^2&\stackrel{\Sigma}{=}&
\frac{ \varepsilon}{4 \pi R^2} \frac{dM(u)}{du} , \label{omega=phi} \\
p+\Omega^2+\Pi^{}_{\mu\nu} n^\mu n^\nu - \frac{\psi^2}{8 \pi}
&\stackrel{\Sigma}{=} &
\frac{ \varepsilon}{4 \pi R^2} \frac{dM(u)}{du} -
\frac {\bar{Q}^2}{8 \pi \bar{R}^4}, \label{p+omega+pill=phi}
\end{eqnarray}
where we have used (\ref{TI}) and (\ref{TEX}).
Combining the last two equations and using (\ref{q=Q}) we get
\begin{equation}
p+\Pi^{}_{\mu\nu}n^\mu n^\nu\stackrel{\Sigma}{=}0 \ ,
\label{p+pill=0}
\end{equation}
which {\em does not involve} any quantity from the space-time
$\bar{\cal V}$.

The equalities (\ref{q=Q}), (\ref{omega=phi}) and (\ref{p+pill=0})
are the main physical equations which relate quantities at both
sides of $\Sigma$. The equation (\ref{omega=phi}) expresses the
fact that the radiated energy density has to be continuous through
the matching hypersurface. On the other hand,
it is easy to check that (\ref{p+pill=0}) is strictly equivalent
to the matching condition (\ref{jc4bis}) (or to (\ref{mc4bis})),
see \cite{fjlls2}. Its physical interpretation is clear,
(\ref{p+pill=0}) says that the total normal pressure has to vanish on the
matching hypersurface. All in all,
we can reformulate the results from this and the previous sections
in the following way
\begin{teorem} \label{teorema1}
Every spherically symmetric metric can be locally matched to a
Vaidya-Reissner-Nordstr\"{o}m solution across
any time-like hypersurface with the properties:
1) the total normal pressure  of
the fluid in the decomposition (\ref{TI}) vanishes on it; and 2)
the total charge enclosed in it is constant.
\end{teorem}
This is a very satisfactory result, which
generalizes previous results for the uncharged case, in particular
those obtained by Misner \& Sharp \cite{MSh} and Bel \& Hamoui \cite{Bel}
for the case with a Schwarzschild
exterior and absence of radiation; and those of Fayos et al. \cite{fjlls2}
for the Vaidya's radiative solution. It is also a generalization
of previous results obtained by Oliveira \& Santos \cite{Oliveira}
for charged shear-free fluids, although this case is better adapted
to the second interpretation given in (\ref{TI2}). We can actually
reformulate the previous theorem in this case as follows:
\begin{teorem}\label{teorema2}
Every spherically symmetric metric can be locally matched to a
Vaidya-Reissner-Nordstr\"{o}m solution across any
time-like hypersurface whose total normal
pressure in the decomposition (\ref{TI2}) equals the heat flux on it,
and such that the total
charge enclosed inside it is constant.
\end{teorem}

\subsection{Brief note on the existence of a surface charge on $\Sigma$}
If we allowed for the existence of infinite jumps in the electromagnetic
4-current the situation would be slightly different, and for example
condition (\ref{p+pill=0}) would not be valid. Instead, we would
have, from (\ref{omega=phi}) and (\ref{p+omega+pill=phi}):
\begin{displaymath}
p+\Pi^{}_{\mu\nu} n^\mu n^\nu - \frac{Q^2}{8 \pi R^4}
\stackrel{\Sigma}{=} -\frac {\bar{Q}^2}{8 \pi \bar{R}^4} .
\end{displaymath}
This can be easily understood by noting that
the normal tensions exerted by the
electromagnetic fields on $\Sigma$, given by
$-Q^2/(8 \pi R^4)$ and $-\bar{Q}^2/(8 \pi\bar{R}^4)$,
do not compensate reciprocally. A non-zero
``normal pressure exerted by the fluid'' is thus needed in order to
satisfy the equality of {\em total} normal pressures as required by
(\ref{Tsigmacn}). Given this, we can generalize
theorems \ref{teorema1} and \ref{teorema2} as
\begin{teorem}\label{teorema3}
If infinite jumps in the electromagnetic 4-current
are allowed, every spherically symmetric
metric can be locally matched to the Vaidya-Reissner-Nordstr\"{o}m
solution across any time-like hypersurface $\Sigma$ such that
\begin{displaymath}
4\pi R^4 (p+ \Pi_{\mu\nu} n^\mu n^\nu)- Q^2 \stackrel{\Sigma}{=}
constant \leq 0\ .
\end{displaymath}
\end{teorem}
\begin{teorem}\label{teorema4}
If we allow for infinite jumps in the electromagnetic 4-current,
every spherically symmetric metric can be
locally matched to a Vaidya-Reissner-Nordstr\"{o}m solution
across any time-like hypersurface satisfying
\begin{displaymath}
4 \pi R^4 (\tilde{p} + \tilde{\Pi}_{\mu\nu} n^\mu n^\nu - h_\alpha
n^\alpha) -Q^2 \stackrel{\Sigma}{=} constant \leq 0\ .
\end{displaymath}
\end{teorem}
Clearly these theorems are less restrictive than the
corresponding theorems in the absence of surface charge, since the
time-like hypersurfaces satisfying the requirements in theorem \ref{teorema1}
(or \ref{teorema2})  will also satisfy the corresponding requirements
in theorem \ref{teorema3} (or, respectively, \ref{teorema4}),
but not the other way round.

\section{Other important physical consequences}\label{secemc}
In this section we are going to prove that the energy and matching
conditions imply several important physical consequences, namely,
(i) inequalities valid on the matching hypersurface, one of them
involving only quantities of $\cal V$; (ii) the non-spacelike
character of the matching hypersurface under general conditions;
and (iii) several limits on the total charge of the model.

To that end, we are going to exploit the energy conditions on
${\bf P}$, given by (\ref{ec1o}), (\ref{ec3o}), and
(\ref{ec2}-\ref{ec5}) supplemented with the fact that $\phi=0$
and, therefore, $\Phi^2=\psi^2$. This group of conditions
constrain the functions $m(u,R)$, $\beta(u,R)$ and $\psi(u,R)$
that we can choose for ${\cal V}$ and, when specialized to
$\Sigma$, give physical restrictions valid on the matching
hypersurface. For instance, the matching conditions for the
$\varepsilon \bar{\varepsilon}=\epsilon_n$ case\footnote{Obviously
the $\varepsilon \bar{\varepsilon}=-\epsilon_n$ case can be
treated similarly.} are the relations (\ref{jc1})-(\ref{jc3}) and
(\ref{jc4bis}). We can isolate $m_{,R}$ and $m_{,u}$ from them in
terms of magnitudes from $\bar{\cal{V}}$
\begin{eqnarray}
m_{,R} &\stackrel{\Sigma}{=}& 2 \bar{R} \left( (\bar{\chi} -
\bar{\varepsilon}\bar{R}') \beta_{,R} + \frac{\bar{Q}^2}{4
\bar{R}^3}\right)\\
\varepsilon m_{,u} &\stackrel{\Sigma}{=}&\bar{\varepsilon} e^{2
\beta} \left( \bar{M}_{,\bar{u}} - 2 \bar{R} \beta_{,R} (\bar{\chi}
-\bar{\varepsilon}\bar{R}') \bar{R}'\right)
\end{eqnarray}
where $\bar{R}'\equiv \dot{\bar{R}}/\dot{\bar{u}}$. Combining these
with  (\ref{ec2}) and (\ref{ec3o}) we arrive at
\begin{eqnarray}
0&\stackrel{\Sigma}{\leq}& \bar{R} \beta_{,R}(\bar{\chi} - 2
\bar{\varepsilon}\bar{R}') \label{df1}\\
0&\leq&
\bar{\varepsilon}\bar{M}_{,\bar{u}}\stackrel{\Sigma}{\leq}\frac{1}{2}\bar{R}
\beta_{,R}(\bar{\chi} - 2 \bar{\varepsilon}\bar{R}')^2 \label{df2}
\end{eqnarray}
where use have been made of the energy condition (\ref{cevb}).
On the other hand, if we differentiate (\ref{jc3}) on $\Sigma$ and replace
$\dot{R}/\dot{u}$ with the value arising from (\ref{jc4bis}) we find
\begin{equation}
\varepsilon m_{,u} \beta_{,R} \stackrel{\Sigma}{\geq}\frac{ e^{2 \beta}}{ 8
R^6} ( Q^2 - 2 R^2 m_{,R})(Q^2 - 2 R^2 m_{,R} + 4 \chi
\beta_{,R} R^3).\label{df3}
\end{equation}
where again the fulfillment of (\ref{cevb}) has been used. Unlike
(\ref{df1}) and (\ref{df2}), this inequality (\ref{df3}) only
relates quantities from $\cal{V}$ on $\sigma$, and it will be
important in the construction of models. On the other hand,
inequalities (\ref{df1}), (\ref{df2}) and (\ref{df3}) use up all
the reciprocal implications of the dominant energy condition at
both sides of the matching hypersurface.

More interestingly, we can derive under which circumstances the
matching hypersurface will be actually non-spacelike\footnote{Here
we will use that (\ref{jc4bis}) and (\ref{mc4bis}) are valid no
matter what the character of the matching hypersurface is, as was
shown in \cite{Merce}.}. The condition for this is $\dot{u}(2
\varepsilon \dot{R}-e^{2\beta} \chi \dot{u}) \leq 0$. We can
consider two possibilities:

\begin{enumerate}
\item $\dot{u}=0$, so that
$\Sigma$ is null. Then we get:
\begin{itemize}
\item Case $\varepsilon\bar{\varepsilon}=\epsilon_n$, from (\ref{jc4bis}):
\begin{displaymath}
\beta,_R \stackrel{\Sigma}{=} 0
\end{displaymath}
\item Case $\varepsilon\bar{\varepsilon}=-\epsilon_n$, from (\ref{mc4bis}):
\begin{displaymath}
\frac{m,_R}{R}-\chi \beta,_R -\frac{Q^2}{2 R^3}
\stackrel{\Sigma}{=} 0
\end{displaymath}
\end{itemize}

\item $\dot{u}>0$, so that
\begin{equation}
\varepsilon \frac{\dot R}{\dot u} \stackrel{\Sigma}{\leq}
\frac{e^{2\beta}}{2}\chi \,  .\label{tlcs}
\end{equation}
Isolating $\dot{R}/\dot{u}$ from (\ref{jc4bis}) [or from
(\ref{mc4bis}) for the second case] and replacing it into
(\ref{tlcs}) we get
\begin{itemize}
\item Case $\varepsilon\bar{\varepsilon}=\epsilon_n$:
\begin{equation}
\left(\mbox{if}\ \ \beta,_R \stackrel{\Sigma}{\neq} 0 \right)\
\Rightarrow \ \ \frac{m,_R}{R} - {\chi} \beta,_R -\frac{Q^2}{2
R^3} \stackrel{\Sigma}{\geq}0 \label{tls1}
\end{equation}
\item Case $\varepsilon \bar{\varepsilon}=-\epsilon_n$:
\begin{equation}
\left(\mbox{if}\ \ \frac{m,_R}{R}-\chi \beta,_R -\frac{Q^2}{2 R^3}
\stackrel{\Sigma}{\neq} 0\right)\ \Rightarrow \ \  \chi^2
\beta,_R+\epsilon \frac{2 m,_u e^{-2\beta}}{R}
\stackrel{\Sigma}{\geq}0 \label{tls2}
\end{equation}
\end{itemize}
\end{enumerate}
These relations are satisfied if (\ref{ec2}) and
(\ref{ec3o}) hold. Furthermore, $\beta_{,R} >0$ and $\chi^2
\beta,_R+\epsilon (2 m,_u e^{-2\beta})/R > 0$, {\em with the strict
inequality}, are part of the energy conditions for energy-momentum
tensors of type I (see \cite{HE} for definitions).
Therefore, we have
\begin{propos}\label{temppreli}
If the energy-momentum tensor in $\cal{V}$ is of ``type I'' and
its non-electromagnetic part satisfies the dominant energy
conditions, and if no infinite jumps are allowed for the
electromagnetic 4-current, then any matching hypersurface $\Sigma$
with a Vaidya-Reissner-Nordstr\"om solution is necessarily
non-spacelike.
\end{propos}
More specifically, assuming that (\ref{ec2}) holds, it is easy to
see that the matching hypersurface will generically be timelike,
and it may be null only in the very particular case that
$m,_R/R-\chi \beta,_R -Q^2/(2 R^3) \stackrel{\Sigma}{=} 0$ holds.

On the other hand, \textit{type II} energy-momentum tensors are
usually associated with zero rest mass fields (recall that, for
instance, the Vaidya and V-RN solutions have a \textit{type II}
energy-momentum tensor). We can consider the question of whether
or not a proposition equivalent to \ref{temppreli} (relating the
energy conditions for $\cal V$ with the non-spacelike nature of
$\Sigma$) can be found when the energy-momentum tensor is \textit{type
II}. It is easy to see that such proposition \emph{cannot exist
in general}, since the matching conditions in the previous section
show that, in the particular case of two matchable V-RN
space-times, the nature of the matching hypersurface can be
arbitrary independently of any other conditions.

Let us consider finally the important question of whether there
are any upper limits on the total charge of the model. From
proposition \ref{propm>0} evaluated on $\Sigma$ and using
(\ref{jc3}) [or (\ref{mc3})] we immediately obtain
\begin{propos}\label{limcarga}
If $m(u,0)\geq 0$ and the dominant energy conditions hold, then
\begin{equation}
\bar{Q}^2\stackrel{\Sigma}{\leq} 2 \bar{M}(\bar{u})
\bar{R}(\bar{u}). \label{Qmax}
\end{equation}
\end{propos}
This is a key result in our treatment, for it can be interpreted
as a constraint on the charge enclosed in $\Sigma$. Notice that
this constraint exists even if there are curvature singularities
at $R=0$ provided that $m(u,0)\geq 0$. Ponce de Le\'on \cite{PdL}
(see also \cite{Bonnor}) obtained a similar result for the much
more restricted case of {\em static, singularity-free} space-times
with a charged {\it perfect fluid} matched to a {\em pure}
Reissner-Nordstr\"om exterior. A general result for the general
class of static singularity-free metrics
can also be deduced from the limits found in \cite{MMPS}, and a more recent work
is presented in \cite{Mak}. The fact of considering only static
space-times is, possibly, the most important restriction in the above-mentioned
papers, since then the results can only be applied
to ``limiting configurations'' for charged static spheres.
Only our general approach allows for an interpretation of
(\ref{Qmax}) in terms of the dynamics of the problem and allows
for a general analysis of the possible existence of \textit{naked
singularities} (see e.g. \cite{Penrose}\cite{Joshi} for
definitions and relevance) and their creation.

With regard to this, observe that (\ref{Qmax}) can be also viewed in other
interesting ways when expressed as
$$
\bar{R}_\Sigma(\bar{u})\geq\frac{\bar{Q}^2}{2\bar{M}(\bar{u})}.
$$
This inequality implies that any space-time $\cal V$ providing a
physically realistic model for the interior of a collapsing star
either has its limit surface at values of
$\bar{R}_\Sigma(\bar{u})\geq \bar{Q}^2/(2\bar{M}(\bar{u}))$, or
otherwise a {\em necessarily timelike} singularity\footnote{The
singularity must be timelike because $m(u,0)<0$.} (therefore
locally naked) must develop at $R=0$. The question on whether this
radius can effectively be reached lies beyond the purposes of this
article, but you can consult \cite{Cauchy} and references therein
for an interesting discussion at this respect.

There are also constraints on the charge $Q$ of $\cal V$ due to the
the fulfillment of DEC. One of them is given by (\ref{df3}). Apart from this,
inequalities (\ref{ec2}) and (\ref{ec4})
define two new possible maximum values for $Q$: the first one is just
the appropriate specialization of (\ref{qm})
\begin{equation}
Q^2 \leq 2 R^2 (m,_R- R{\chi} \beta,_R) ,\label{Q1}
\end{equation}
while the second one reads
\begin{equation}
Q^2 \leq \frac{1}{2} R^4 (\lambda + P_2) .\label{Q2}
\end{equation}
The righthand sides of these two inequalities vanish at $R=0$
if there is no curvature singularity there, as is clear
from (\ref{ro}). Thus, if $\Sigma$ arrived
at the vicinity of $R=0$ with non-zero charge,
the previous conditions would be violated from a certain value of $R$ on
(decreasingly). This result is similar to that in proposition
\ref{limcarga}: $\Sigma$ can only get arbitrarily close
to $R=0$ if either a curvature singularity already exists,
or develops, there.

\section{Conclusions}
The aim of this work is to provide a theoretical framework
for the construction of global models describing stars and voids
in General Relativity. Our only assumption is, apart from the spherical symmetry
of the spacetime, that the exterior of the star (respectively the
gravitational field within the void) is represented by a
Vaidya-Reissner-Nordstr\"om solution properly matched to an interior
(resp.\ exterior) spacetime with the only restriction that the dominant energy
conditions are fulfilled everywhere. Thus, we are treating the general case
in which the stars/voids may be charged and radiating.

A summary of our main results is the following. To start with, in proposition
\ref{propm>0} we saw that the DEC imply the non-negativity of the mass function
everywhere if $m(u,0)\geq 0$. This result has important consequences in
the rest of the paper, as for instance the restriction which imposes on any
\textit{spherically symmetric} electromagnetic fields, see section \ref{seccem}.
More important consequences are the restrictions on the presence and distribution
of electric charge, first in a neighbourhood of $R=0$ (section \ref{seccem})
and then on the whole spacetime (formulae (\ref{Q1}-\ref{Q2}) and proposition
\ref{limcarga}).

As a specially important consequence, we have shown in full
generality that the {\em total} charge of any physically acceptable spherically
symmetric object has un upper bound related to its size and mass given by
(\ref{Qmax}). Here, by physically acceptable
we mean that the object satisfies $m(u,0)\geq 0$ which in particular includes
all bodies which are regular everywhere. This is a result genuine to General
Relativity for, as is well known, Classical Electrodynamics does not impose any
limitation on the charge that an object can possess. Furthermore, our
result is completely general and does not depend on the assumptions of staticity
used in its previous partial and particular versions found in
\cite{PdL,Bonnor,MMPS} and references therein, or in
the more recent \cite{Mak}. In this last work an inequality relating the mass,
the radius and the charge of general \emph{static} matter distributions has
been found by using the gravitational field equations (in a
similar way as Buchdahl \cite{Buchdahl} obtained the classical
maximum mass-radius ratio for uncharged stable stars). Their result
shows that a charged ($Q<M$) stable star must obey the inequality (\ref{Qmax}).
However, as is clear, our treatment provides the limit
(\ref{Qmax}) in dynamical situations, including evolving voids or the important
case of collapsing, or rebounding, stars.

The appearance of a maximum value for the charge can be alternatively read
as the existence of a minimum value for $R_\Sigma(u)$. This result can be
interpreted, for the case of a charged star --no matter how small its
charge is-- that obeys the DEC as stating that, if the star ever
reaches values of $R$ lower than the minimum radius, then a necessarily
timelike singularity, ergo locally naked, has to develop at $R=0$.

All our results have been derived by properly matching the
interior and exterior spacetimes, and here we have taken into
account the {\em pure electromagnetic} matching conditions as well
as the gravitational ones, which were fully derived in section
\ref{capenlaces} both in the case with no infinite jumps for the
total electromagnetic 4-current
---which agrees with experimental results--- or in the case with them ---this
is the usual approach in the literature but, in our opinion, it is
unrealistic from a physical point of view, except as an
approximation to the real world---. We have also proved that, in
generic situations, there is a 2-parameter family of matching
hypersurfaces which were later shown to be non-spacelike in
Proposition \ref{temppreli}. The general result that the vast
majority of spherically symmetric spacetimes can be matched to a
V-RN solution (or one of its particular cases) has a clear
physical interpretation given by Theorems \ref{teorema1} and
\ref{teorema2}. These results generalize the ones obtained in
\cite{fjlls2,MSh,Bel,Oliveira} for several particular cases
arising if there is no charge or radiation. However, let us remark
that even if the {\em total} charge of the modelized object is
zero and, consequently, the metric for the space-time $\bar{\cal
V}$ is the Vaidya metric (or its specializations), still a radial
distribution of charge is allowed in ${\cal V}$, even a
time-dependent one, as long as the total charge is zero. This fact
allows for a possible description of the charge redistribution
phenomena which could take place in the interior of a star. This
is the case of, for instance, the {\it hybrid stars}, in which,
despite global charge neutrality, the local non-neutrality is
energetically favoured (see \cite{Glendenning}).

In our next article (\textit{Spherically symmetric models for
charged radiating stars and voids II: Practical approach}) we
analyze in depth some of the theoretical formulations proposed in
this work and examine some models for stars and voids under the
scope of our approach.

\section*{Acknowledgements}
J.M.M.S. is grateful to the EPSRC grant MTH 03 RAJC6 for financial
help for a 3-month visit to Queen Mary, University of London,
where this work was partly carried out. F.F. acknowledges
financial support from the D.G.R. of the Generalitat de Catalunya
(grant 2000SGR/23) and the Spanish Ministry of Education (contract
2000-0604).

\end{document}